\documentclass[preprint1]{aastex}

\newcommand{\be}{\begin{equation}}
\newcommand{\ee}{\end{equation}} 

\shorttitle{Diffusive Acceleration}
\shortauthors{Fatuzzo and Melia}

\begin{document}

\title{Fitting the Union2.1 SN Sample with the $R_{\rm h}=ct$ Universe}

\author{F. Melia\thanks{John Woodruff Simpson Fellow.}}
\affil{Department of Physics, The Applied Math Program, and Steward Observatory, \\ 
The University of Arizona, AZ 85721}
\email{fmelia@email.arizona.edu}

\begin{abstract}
The analysis of Type Ia supernova data over the past decade has been a notable 
success story in cosmology. These standard candles offer us an unparalleled 
opportunity of studying the cosmological expansion out to a redshift of $\sim 1.5$. 
The consensus today appears to be that $\Lambda$CDM offers the best explanation 
for the luminosity-distance relationship seen in these events. However, a significant 
incompatibility is now emerging between the standard model and other equally 
important observations, such as those of the cosmic microwave background. 
$\Lambda$CDM does not provide an accurate representation 
of the cosmological expansion at high redshifts ($z>>2$). It is 
therefore essential to re-analyze the Type Ia supernova data in light of the 
cosmology (the $R_{\rm h}=ct$ Universe) that best represents the Universe's 
dynamical evolution at early times. In this paper, we directly compare the 
distance-relationship in $\Lambda$CDM with that predicted by $R_{\rm h}=ct$, and 
each with the Union2.1 sample, and show that the two theories produce virtually 
indistinguishable profiles, though the fit with $R_{\rm h}=ct$ has not yet been
optimized. This is because the data cannot be determined independently of the 
assumed cosmology---the supernova luminosities must be evaluated by optimizing 
4 parameters simultaneously with those in the adopted model. This renders the data 
compliant to the underlying theory, so the model-dependent data reduction should
not be ignored in any comparative analysis between competing cosmologies. In this 
paper, we use $R_{\rm h}=ct$ to fit the data reduced with $\Lambda$CDM, and
though quite promising, the match is not perfect. An even better fit would
result with an optimization of the data using $R_{\rm h}=ct$ from the beginning.
\end{abstract}


\section{Introduction}
We now have several methods at our disposal for probing the expansion history of the
Universe, but none better than the use of Type Ia supernovae (Perlmutter et 
al. 1998, 1999; Garnavich et al. 1998; Schmidt et al. 1998; Riess et al. 1998). Producing
a relatively well-known luminosity, these events function as reasonable standard candles, 
under the assumption that the power of both near and distant events can be 
standardized with the same luminosity versus color and light-curve shape relationships.

Over the past decade, the combined efforts of several groups have led to an
impressive accumulation of events and an ever improving quality of individual
measurements. More recently, Kowalski et al. (2008) devised a framework for
analyzing such data sets in a homogeneous manner, creating an evolving
compilation currently known as the Union2.1 sample (see Suzuki et al. 2012
for its most recent incarnation), which contains 580 supernova detections.
At the highest redshifts ($z>1$), the Hubble Space Telescope plays a key 
role, not only in discovering new events, but also in providing high-precision 
optical and near-IR lightcurves (see, e.g., Riess et al. 2004; Kuznetsova et 
al. 2008; Dawson et al. 2009). While at lower redshifts, extensive surveys 
from the ground continue to grow the sample at a sustainably high rate (see
references cited in Suzuki et al. 2012).

These events constrain cosmological parameters through a comparison of
their apparent luminosities with those predicted by models over
a range of redshifts. The key point of this exercise is that models with
different expansion histories predict specific (and presumably distinguishable)
distance vs. redshift relationships, which one may match to the observations 
for a comparative analysis.  The consensus today appears to be that $\Lambda$CDM
offers the best explanation for the redshift-luminosity distribution seen in these 
events, and observational work is now focused primarily on refining the fits to
improve the precision with which the model parameters are determined.

When these efforts are viewed through the prism of a more comprehensive cosmological study, 
however, the situation is not quite so simple. The reason is that although $\Lambda$CDM
does very well in accounting for the properties of Type Ia supernovae, it does poorly 
in attempts to explain several other equally important observations, particularly the angular 
distribution of anisotropies in the cosmic microwave background (CMB) and the clustering 
of matter. For example, the angular correlation function of the CMB is so different from 
what $\Lambda$CDM predicts that the probability of it providing the correct expansion 
history of the Universe near recombination is less than $\sim$$0.03\%$ 
(Copi et al. 2009; Melia 2012b). And insofar as the matter distribution
is concerned, $\Lambda$CDM predicts a distribution profile that changes with
scale, seemingly at odds with the observed near universal power-law seen
everywhere below the BAO scale ($\sim 100$ Mpc), leading Watson et al. 
(2011) to categorize it as a ``cosmic coincidence." 

This disparity is the principal topic we wish to 
address in this paper, with a particular emphasis on what the $R_{\rm h}=ct$ 
Universe has to say about the universal expansion implied by the Type Ia
supernova data. As we have shown in earlier papers (see, e.g., Melia 2007,
2012a, 2012b, 2012c, 2012d; Melia \& Shevchuk 2012), and summarized in
\S 2 below, the $R_{\rm h}=ct$ Universe is more effective than $\Lambda$CDM 
in accounting for the properties of the CMB. This raises the question of whether
$\Lambda$CDM is really revealing something different about the local 
Universe---as opposed to what happened much earlier, closer to the time 
of last scattering---or whether it is possible for us to understand how and why
it may simply be mimicking the expansion history suggested by the $R_{\rm h}=ct$ 
Universe at low redshifts.

This is not an easy question to address, principally because of the enormous
amount of work that goes into first establishing the supernova magnitudes,
and then carefully fitting the data using the comprehensive set of parameters
available to $\Lambda$CDM. As we shall see in \S 3 below, part of the problem 
is that the distance moduli themselves cannot be determined completely independently
of the assumed model. This leads to the unpalatable situation in which the data
tend to be somewhat compliant, producing slightly better fits for the input model
than might otherwise occur if they were completely model independent. But
even with these complications, we will show below that $\Lambda$CDM
appears to fit the supernova data well because its seven free parameters 
allow it to relax to the $R_{\rm h}=ct$ Universe at low redshifts. In other
words, we suggest that fits to the Union2.1 sample cannot, by themselves,
distinguish between $\Lambda$CDM and the $R_{\rm h}=ct$ Universe. 
The preponderance of evidence therefore rests with the CMB, which appears to
favor the $R_{\rm h}=ct$ cosmology.

\section{The $R_{\rm h}=ct$ Universe}
The $R_{\rm h}=ct$ Universe is an FRW cosmology that adheres very closely to the restrictions 
imposed on the theory by the Cosmological Principle and Weyl's postulate (Melia 2007; Melia \& 
Shevchuk 2012). Most people realize that adopting the Cosmological principle means we assume 
the Universe is homogeneous and isotropic. But many don't know that this is only a statement
about the structure of the Universe at any given cosmic time $t$. In order to complete the
rationale for the FRW form of the metric, one also needs to know how the time slices at
different $t$ relate to each other. In other words, one needs to know whether or not
the Cosmological principle is maintained from one era to the next. And this is where 
Weyl's postulate becomes essential, for it introduces the equally important assumption
about how the various worldlines propagate forwards in time. Weyl's postulate says
that no two worldlines in the Hubble flow ever cross. Small, local crossings are, of
course, always possible, and we understand that these are due to small departures
from an overall Hubble expansion. But on large scales, the assumption that worldlines
never cross places a severe restriction on how the distance between any two points
can change with time. This is the origin of the relationship between the so-called
proper distance $R(t)$ in FRW cosmologies and the universal expansion factor
$a(t)$, such that $R(t)=a(t)r$, in terms of the unchanging co-moving distance $r$.

But what has not been recognized until recently is that when these two basic tenets 
are taken seriously, they force the gravitational horizon $R_{\rm h}$ (more commonly 
known as the Hubble radius) to always equal $ct$. (The proof of this is straightforward
and may be found, e.g., in Melia \& Shevchuk 2012, and in the more pedagogical
treatment of Melia 2012a.) It is not difficult to see that this equality in turn forces the 
expansion rate to be constant, so the expansion factor $a(t)$ appearing in the Friedmann 
equations  must be $t/t_0$ (utilizing the convention that $a(t_0)=1$ today), where 
$t_0$ is the current age of the Universe.

Over the past several decades, $\Lambda$CDM has developed into a comprehensive 
description of nature, which nonetheless is not entirely consistent with this approach.
Instead, $\Lambda$CDM assumes a set of primary constituents (radiation, matter, and an
unspecified dark energy), and adopts a partitioning among these components that is not theoretically
well motivated. $\Lambda$CDM is therefore an empirical cosmology, deriving many
of its traits from observations. The problem with this, however, is that the observations
are quite varied and cover disparate properties of the Universe, some early in its
history (as seen in the fossilized CMB) and others more recently (such as the formation
of large-scale structure and the aforementioned distribution of Type Ia supernovae). 

In recent work, we have demonstrated why this approach may be useful in establishing
a basis for theoretical cosmology, but why in the end $\Lambda$CDM essentially remains
an approximation to the more precise theory embodied in the $R_{\rm h}=ct$ Universe.  
What is lacking in $\Lambda$CDM is the overall equation of state, $p=-\rho/3$, linking the 
total pressure $p$ and the total energy density $\rho$. This simple relationship is required when
the Cosmological principle and the Weyl postulate are adopted together. In consequence,
$\Lambda$CDM may ``fit" the data in certain restricted redshift ranges, but it fails in other
limits. For example, $R_{\rm h}$ in $\Lambda$CDM fluctuates about the mean it would 
otherwise always have (since it must always be equal to $ct$), leading to the awkward 
situation in which the value of $R_{\rm h}(t_0)$ is equal to $ct_0$ today, but in order 
to achieve this ``coincidence", the Universe had to decelerate early on, followed by a 
more recent acceleration that exactly balanced out the effects of its slowing down at 
the beginning. In \S\S~4 and 5, we shall see how this balancing act functions in a
more quantitative manner (particularly when fitting the Type Ia supernova data), 
and we shall actually learn that the cancellation is even more contrived than it 
seems at first blush.

The problems with $\Lambda$CDM begin right at the outset because of its implied
deceleration just after the big bang. This slowing down brings it into direct conflict 
with the near uniformity of the CMB data (see, e.g., Melia 2012a), requiring the 
introduction of an ad hoc inflationary phase to rescue it. As shown in Melia 
(2012b), however, the recent assessment of the observed angular correlation
function $C(\theta)$ in the CMB simply does not support any kind of inflationary 
scenario. We showed in that paper that, whereas $\Lambda$CDM fails to account 
for the observed shape of $C(\theta)$---particularly the absence of any correlation
at angles greater than $\sim$$60^\circ$---the $R_{\rm h}=ct$ Universe explains
where $C(\theta)$ attains its minimum value (at $\theta_{\rm min}$), the correlation 
amplitude $C(\theta_{\rm min})$ at that angle and, most importantly, why there is 
no angular correlation at large angles. The chief ingredient in $\Lambda$CDM
responsible for this failure to account for the properties of $C(\theta)$ is
inflation, which would have expanded all fluctuations to very large scales. 
Instead, the $R_{\rm h}=ct$ Universe did not undergo such an accelerated 
expansion, so its largest fluctuations are limited to the size of the gravitational 
horizon $R_{\rm h}(t_e)$ at the time of last scattering. 

To be completely fair to the standard model, however, we also point out
that in spite of its failure to account for the distribution of large-scale fluctuations
in the CMB, it does very well in explaining the large-$l$ power spectrum associated
with fluctuations on scales below $\sim 1^\circ$. And recent observations provide
an even more convincing confirmation of the large-$l$ spectrum predicted by
$\Lambda$CDM (see, e.g., Hlozek et al. 2012). More work needs to be carried
out to fully understand this disparity. It could be that unlike the Sachs-Wolfe
effect, the physical processes (such as acoustic fluctuations) influencing the
small-scale behavior are much less dependent on the expansion scenario.
There is a hint that this may be happening from the analysis of Scott et al.
(1995), who demonstrated that the same large-$l$ power spectrum can
result from rather different cosmological models.   

The CMB carries
significant weight in determining which of the candidate models accounts
for the universal expansion, and here we have a situation in which 
$\Lambda$CDM cannot explain the near uniformity of the CMB 
across the sky without inflation, yet with inflation it cannot account for 
the anisotropy of the CMB on large scales.

This distinction alone already demonstrates the superiority of the $R_{\rm h}
=ct$ Universe over $\Lambda$CDM, insofar as the interpretation of the 
CMB observations is concerned. But this is only one of several crucial tests 
affirming the conclusion that $\Lambda$CDM is only an approximation to 
the more precise $R_{\rm h}=ct$ Universe. Here are several other reasons:

$\bullet$ The $R_{\rm h}=ct$ Universe explains why it makes sense to infer a Planck 
mass scale in the early Universe by equating the Schwarzschild radius to the 
Compton wavelength. In $\Lambda$CDM there is no justifiable reason
why a delimiting gravitational horizon should be invoked in an otherwise
infinite Universe (see Melia 2012a for a more pedagogical explanation).

$\bullet$ The $R_{\rm h}=ct$ Universe explains why $R_{\rm h}(t_0)
=ct_0$ today (because they are always equal). In $\Lambda$CDM, 
this equality is just one of many coincidences. As we shall see in subsequent 
sections of this paper, this awkwardness will become apparent in how the 
free parameters in $\Lambda$CDM must be manipulated in order to fit 
the Type Ia supernova data (see, e.g., Figure~4).

$\bullet$ The $R_{\rm h}=ct$ Universe explains how opposite sides of the CMB
could have been in equilibrium at the time ($t_e\sim 10^4-10^5$
years) of recombination, without the need to introduce an ad hoc
period of inflation. Inflation may be useful for other reasons, but
it does not appear to be necessary in order to solve a non-existent
``horizon problem" (Melia 2012d).  

$\bullet$ The $R_{\rm h}=ct$ Universe explains why there is no apparent 
length scale below the BAO wavelength ($\sim 100$ Mpc)  in the observed 
matter correlation function. In their exhaustive study, Watson et al. (2011) 
concluded that the observed power-law galaxy correlation function below
the BAO is simply not consistent with the predictions
of $\Lambda$CDM, which requires different clustering profiles of
matter on different spatial scales. These authors suggested, therefore,
that the galaxy correlation function must be a ``cosmic coincidence."
But this is not the case in the $R_{\rm h}=ct$ Universe because
this cosmology does not possess a Jeans length (Melia \& Shevchuk
2012). Since $p=-\rho/3$, the active gravitational mass
($\rho+3p$) in the $R_{\rm h}=ct$ Universe is zero, so fluctuations 
grow as a result of (negative) pressure only, without any delimiting 
spatial scale.

$\bullet$ The fact that $\rho$ is partitioned into $\approx 27\%$ matter and 
$\approx 73\%$ dark energy is a mystery in $\Lambda$CDM. But in 
the $R_{\rm h}=ct$ Universe, it is clear why $\Omega_m$ must be
$\approx 27\%$, because when one forces $\rho$ to have the
specific constituents $\rho_r$, $\rho_m$, and $\rho_\Lambda$,
only the value $\Omega_m\approx 0.27$ will permit the Universe
to evolve in just the right way to satisfy the condition $R_{\rm h}(t_0)
=ct_0$ today. This condition is always satisfied in the $R_{\rm h}=ct$
Universe, but not in $\Lambda$CDM. Yet the observations today must
be consistent with this constraint imposed by the Cosmological principle
and the Weyl postulate, so all the other evolutionary aspects of the
$\Lambda$CDM cosmology must comply with this requirement.   

$\bullet$ The observed near alignment of the CMB quadrupole and octopole
moments is a statistically significant anomaly for $\Lambda$CDM, but
merely lies within statistically reasonable expectations in the 
$R_{\rm h}=ct$ Universe (Melia 2012c). This again has to do with 
the finite fluctuation size, limited by the gravitational horizon 
$R_{\rm h}(t_e)$ at the time of last scattering.

\section{The Union2.1 Supernova Sample}
Let us now briefly review the contents of the Union2.1 sample, and summarize 
the key steps taken during its assembly. As the samples have grown and
become better calibrated, evidence has emerged for a correlation between
host galaxy properties and SN luminosities, after corrections are made for
lightcurve width and SN color (Hicken et al. 2009). For example, Type Ia
supernovae in early-type galaxies appear to be brighter (by  about
$0.14$ mag) than their counterparts in galaxies of later type.  A similar
relationship appears to exist between Hubble residuals and host galaxy
mass (Kelly et al. 2010; Sullivan et al. 2010; 
Lampeitl et al. 2010). Uncorrected, such relationships can lead to 
significant systematic error in determining the best fit cosmology.

Additional sources of uncertainty arise for astrophysical reasons,
including the color correction that must be applied to SN luminosities.
The so-called redder-fainter relation apparently arises from at least
two mechanisms: extinction from interstellar dust, and probably
some intrinsic relation between color and luminosity produced by
the explosion itself or by the surrounding medium. It is difficult to
justify the argument that this redder-fainter relationship should
behave in the same way at all redshifts, but there's little else one
can do because the two effects are very difficult to disentangle 
(see, e.g., Suzuki 2012). 

Combining the many available datasets into a single compilation
(the Union2.1 sample) has obvious advantages, treating all supernovae
on an equal footing and using the same lightcurve fitting, but this
process brings its own set of possible errors, including the fact that
the systematics may be different among the various datasets. As we
shall see shortly, this multitude of uncertainties makes it impossible
to determine the supernova luminosities without some prior assumption
about the underlying cosmological model. 

The procedure for determining each individual Type Ia supernova
luminosity requires a fit to the lightcurve using three parameters
(aside from those arising in the cosmological model itself). These
are (1) an overall normalization, $x_0$, to the time dependent
spectral energy distribution of the supernova, (2) the deviation,
$x_1$, from the average lightcurve shape, and (3) the deviation,
$c$, from the mean Type Ia supernova $B-V$ color.  These three
parameters, along with the assumed host mass, are then combined
to form the distance modulus
\begin{equation}
\mu_B=m_B^{\rm max}+\alpha\cdot x_1-\beta\cdot c
+\delta\cdot P(m_*^{\rm true}<m_*^{\rm threshold})-M_B\;,
\end{equation}
where $m_B^{\rm max}$ is the integrated $B$-band flux at
maximum light, $M_B$ is the absolute $B$-band magnitude
of a Type Ia supernova with $x_1=0$, $c=0$, and $P(m_*^{\rm true}<
m_*^{\rm threshold})=0$. Also, $m_*^{\rm threshold}=10^{10}\,
M_\odot$ is the threshold host-galaxy mass used for the
mass-dependent correction, and $P$ is a probability function 
assigning a probability that the true mass, $m_*^{\rm true}$, is 
less than the threshold value, when an actual mass measurement 
$m_*^{\rm obs}$ is made. 

It is quite evident that the task of accurately determining $\mu_B$
for each individual supernova is arduous indeed. The Supernova
Cosmology Project calls the parameters $\alpha$, $\beta$, $\delta$,
and $M_B$ ``nuisance" parameters, because they cannot be evaluated
independently of the assumed cosmology. They must be fitted simultaneously
with the other cosmological parameters, e.g., emerging from $\Lambda$CDM.

But this is quite a serious problem that cannot be ignored, e.g., when comparing
the overall fits to the data with $\Lambda$CDM and the $R_{\rm h}=ct$ Universe
(or any other cosmology, for that matter) because, as we have already alluded
to above, this procedure makes the data at least somewhat compliant to the
assumed cosmological model. 

\section{Theoretical Fits}
The best fit cosmology is determined by an iterative $\chi^2$-minimization
of the function
\begin{equation}
\chi_{\rm stat}^2=\sum_{\rm supernovae}{\left[\mu_B(\alpha,\beta,\delta,
M_B)-5\log10(d_L(\xi,z)/10\;{\rm pc})\right]^2\over \sigma_{\rm lc}^2+
\sigma_{\rm ext}^2+\sigma_{\rm sample}^2}\;,
\end{equation} 
where $\xi$ stands for all the cosmological parameters that define the fitted
model (with the exception of the Hubble constant $H_0$), $d_L$ is the
luminosity distance, and $\sigma_{\rm lc}$ is the propagated error from the
covariance matrix of the lightcurve fits. The uncertainties due to host
galaxy peculiar velocities, Galactic extinction corrections, and gravitational
lensing are included in $\sigma_{\rm ext}$ (which is evidently of order $0.01$ mag), 
and $\sigma_{\rm sample}$ is a floating dispersion term containing sample-dependent 
systematic errors. Its value is obtained by setting the reduced $\chi^2$ to 
unity for each sample in the compilation. For the Union2.1 catalog, $\sigma_{\rm sample}$
is approximately $0.15$ mag (Suzuki et al. 2012).

The luminosity distance is defined by the relation
\begin{equation}
d_L\equiv a(t_0)r_e{a(t_0)\over a(t_e)}\;,
\end{equation}
where $a(t_0)$ is the expansion factor at the present cosmic time $t_0$, $r_e$ is 
the comoving distance to the source, and $t_e$ is the time at which the source at 
$r_e$ emitted the light we see today. In $\Lambda$CDM, the density is comprised 
of three primary components,
\begin{equation}
\rho=\rho_r+\rho_m+\rho_{\Lambda}
\end{equation}
where, following convention, $\rho_r$ is the density due to radiation,
$\rho_m$ is the matter density and $\rho_\Lambda$ represents dark
energy, usually assumed to be a cosmological constant $\Lambda$. 
Dividing through by the critical density today,
\begin{equation}
\rho_c\equiv {3c^2H_0^2\over 8\pi G}\;,
\end{equation}
we may write
\begin{equation}
\Omega\equiv {\rho\over\rho_c}=\Omega_r+\Omega_m+\Omega_\Lambda
\end{equation}
where, in obvious notation, $\Omega_r\equiv \rho_r/\rho_c$, $\Omega_m\equiv
\rho_m/\rho_c$, and $\Omega_\Lambda\equiv \rho_\Lambda/\rho_c$.

\begin{figure}
\figurenum{1}
\begin{center}
{\centerline{\epsscale{0.90} \plotone{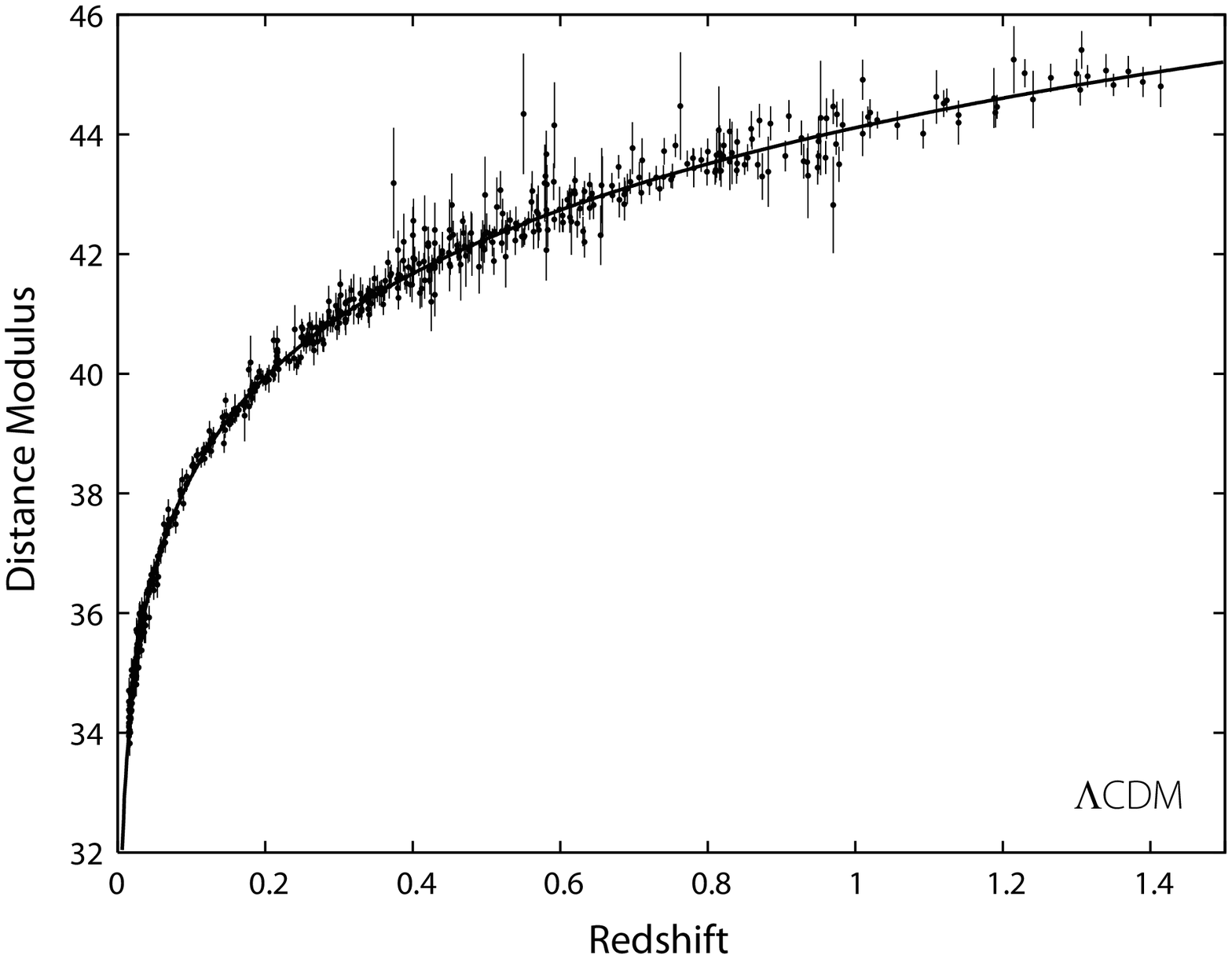} }}
\end{center}
\figcaption{Hubble diagram for the Union2.1 compilation (of 580 Type Ia 
supernova events). The solid curve
represents the best-fitted $\Lambda$CDM model for a flat universe
and matter energy density $\Omega_m=0.27$ (see text). (Adapted from Suzuki et al. 2012)}
\end{figure}

In a spatially flat universe ($k=0$), $\Omega=1$, so if we assume that
$\rho_r\sim a^{-4}$, $\rho_m\sim a^{-3}$, and $\rho_\Lambda=$ constant
then, from the geodesic equation,
\begin{equation}
c\,dt=-a(t)\,dr\;,
\end{equation}
we get
\begin{equation}
r_e={c\over H_0}{1\over a_0^2}\int_{a_e}^{a_0}
{da\over\sqrt{\Omega_r+a\Omega_m/a_0+(a/a_0)^4\Omega_\Lambda}}\;,
\end{equation}
where $a_0\equiv a(t_0)$ and $a_e\equiv a(t_e)$. And changing the variable
of integration to $u\equiv a/a_0$, we find that in $\Lambda$CDM
\begin{equation}
d_L^{\Lambda{\rm CDM}}={c\over H_0}(1+z)\int_{1\over 1+z}^1
{du\over\sqrt{\Omega_r+u\Omega_m+u^4\Omega_\Lambda}}\;.
\end{equation}

Figure~1 shows the Union2.1 sample of 580 Type Ia supernovae, along with
the best fit $\Lambda$CDM model, in which dark energy is a cosmological
constant, $\Lambda$, its equation-of-state parameter is $w=-1$, and
$\Omega_m=0.27$. As is well known by now, the theoretical fit is excellent.
In addition, for these best-fit parameters, the expansion of the Universe
switched from deceleration to acceleration at $z\approx 0.75$, which corresponds
to a look-back time (in the context of $\Lambda$CDM) of $\approx 6.6$ Gyr, roughly
half the current age of the Universe. Equality between $\rho_m$ and $\rho_\Lambda$
occurred later, at $z\approx 0.39$, corresponding to a look-back time of about 
$4.2$ Gyr. 

However, the purpose of this paper is not so much to demonstrate this well known result 
but, rather, to see if our suggested resolution of the disparity between the predictions of 
$\Lambda$CDM and the CMB observations is also consistent with the cosmological 
expansion implied by the Type Ia supernova data. Since the $R_{\rm h}=ct$ Universe is 
superior to $\Lambda$CDM in at least some fitting of the CMB data, the question now is whether it 
can also account for the measured expansion at lower redshifts.

For the $R_{\rm h}=ct$ Universe, we have
\begin{equation}
{a(t_0)\over a(t_e)}={t_0\over t_e}\;,
\end{equation}
from which we get
\begin{equation}
r_e=\ln\left({t_0\over t_e}\right)\;.
\end{equation}
And with
\begin{equation}
1+z={a(t_0)\over a(t_e)}\;,
\end{equation}
it is easy to see that
\begin{equation}
d_L^{R_{\rm h}=ct}={c\over H_0}(1+z)\ln(1+z)
\end{equation}
(where we have also used the fact that $t_0=1/H_0$).

\begin{figure}
\figurenum{2}
\begin{center}
{\centerline{\epsscale{0.90} \plotone{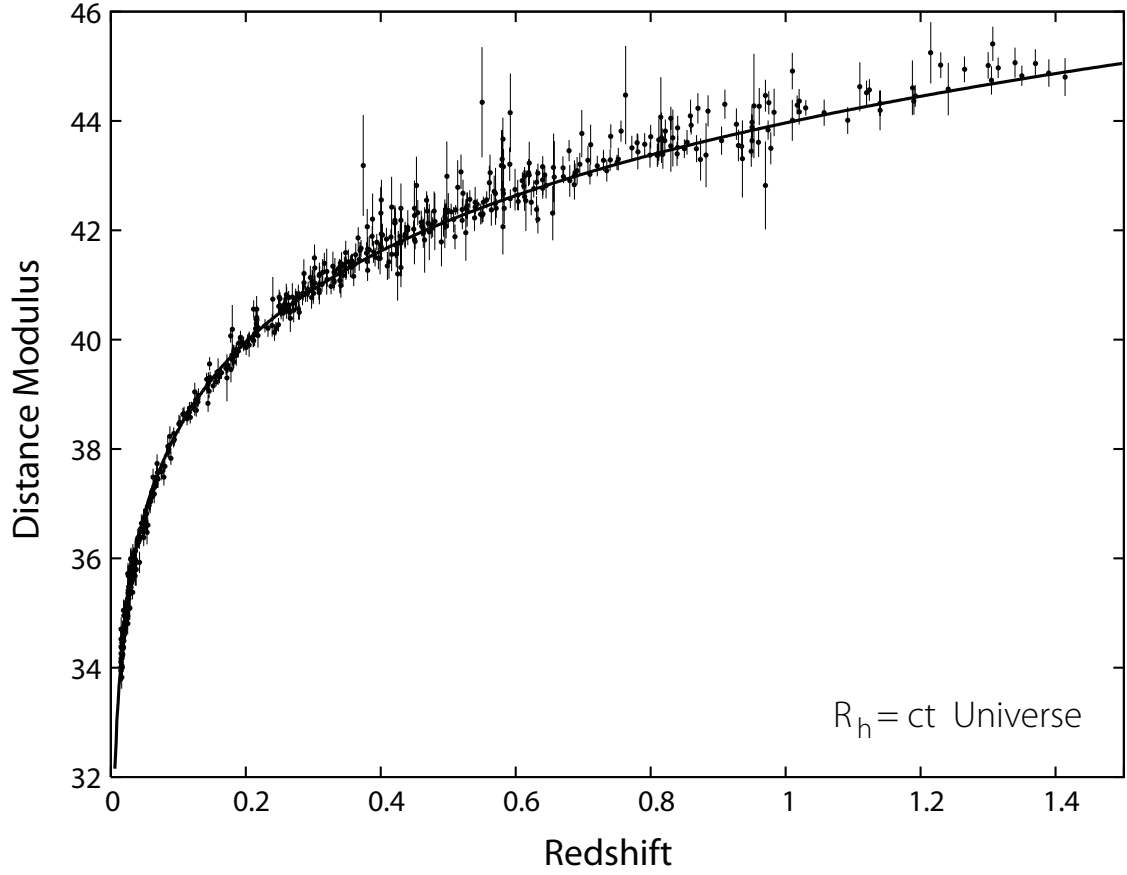} }}
\end{center}
\figcaption{Same as Figure~1, except now the solid curve represents
the $R_{\rm h}=ct$ Universe. (Data are from Suzuki et al. 2012)}
\end{figure}

In carrying out a fit to the data, we note that the chosen value of the Hubble constant 
$H_0$ is not independent of $M_B$. In other words, one can vary either of these parameters,
but not both separately. Therefore, if we allow $M_B$ to remain a member of the set of
variables optimized to find the ``best-fit" supernova luminosities (as described in \S~3 above),
then the $R_{\rm h}=ct$ Universe has no free parameters. In contrast, $\Lambda$CDM has
several, depending on how one chooses to treat dark energy. These include the value of
$\Omega_m$ and possibly the equation-of-state parameter $w$, if dark energy is not a pure
cosmological constant. In $\Lambda$CDM, these parameters affect the location of the
Universe's transition from deceleration to acceleration, and the location where 
$\rho_m=\rho_\Lambda$.  

It is beyond the scope of this paper to carry out a complete best-fit minimization of $\chi_{\rm stat}^2$
for the $R_{\rm h}=ct$ Universe,
which would require an optimization of all the parameters, including $\alpha$, $\beta$, and $\delta$, among
others. Instead, we will adopt the parameters identified by the Supernova Cosmology Project as the
best-fit values for the $\Lambda$CDM model shown in Figure~1, save for one variable---the absolute
magnitude $M_B$. In other words, we will still determine a best fit to the data using the $R_{\rm h}=ct$ 
cosmology, but only by minimizing $\chi_{\rm stat}^2$ with respect to $M_B$. This is not ideal,
of course, because the fit would be even better, were we to allow all of the available parameters to vary,
but even this very simplified approach is sufficient to prove our point and it is easy to digest, so it should
suffice for now.

\begin{figure}
\figurenum{3}
\begin{center}
{\centerline{\epsscale{0.90} \plotone{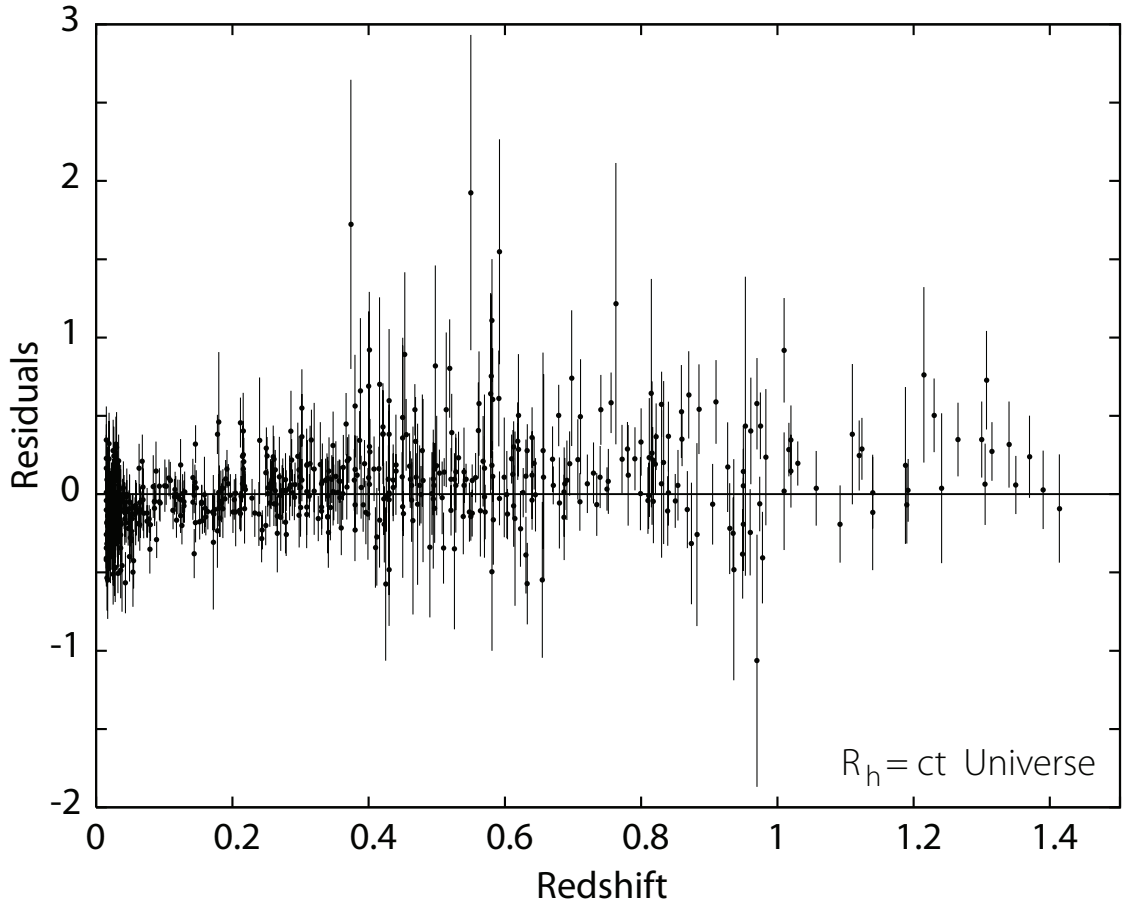} }}
\end{center}
\figcaption{Hubble diagram residuals where the $R_{\rm h}=ct$
Universe profile has been subtracted from the Union2.1 distance
moduli. (Data are from Suzuki et al. 2012)}
\end{figure}

The $R_{\rm h}=ct$ distance profile is compared to the Union2.1 data in Figure~2. The optimization
procedure we have just described results in a ``best-fit" value for $M_B$ that is $0.126$ mag greater 
than that obtained with $\Lambda$CDM. However, for this illustration, all the other parameters are identical
to those calculated for $\Lambda$CDM (resulting in the fit shown in Figure~1). For completeness, we
also show the Hubble diagram residuals corresponding to the $R_{\rm h}=ct$ cosmology in Figure~3.

\section{Discussion}
The fits are so similar that even if there weren't any direct relationship between the two cosmologies,
one would have to suspect that there exists yet another ``cosmic coincidence" in $\Lambda$CDM.
But we will now argue that in fact this is not a coincidence---that $\Lambda$CDM is merely relaxing
to the $R_{\rm h}=ct$ Universe when all of its parameters are optimized to fit the data. To help
with this discussion, let us look at the two distance vs. redshift profiles side by side in Figure~4.

\begin{figure}
\figurenum{4}
\begin{center}
{\centerline{\epsscale{0.90} \plotone{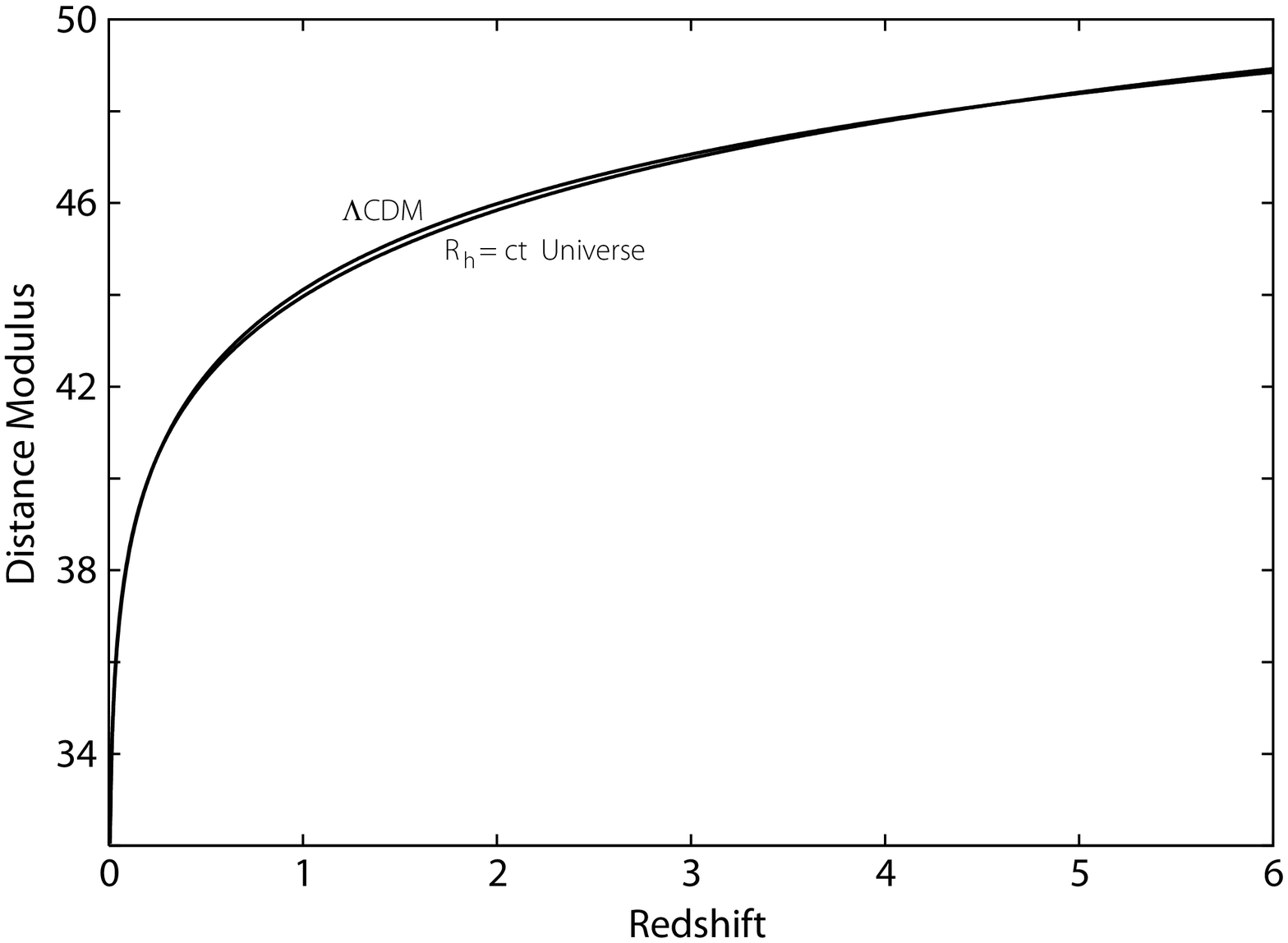} }}
\end{center}
\figcaption{A side-by-side comparison of the best fitted $\Lambda$CDM
and $R_{\rm h}=ct$ Universe cosmologies shown in Figures~1 and 2,
except here extended to a much larger redshift, $z\rightarrow 6$.
The most noticeable feature of this comparison is how closely the
best-fitted $\Lambda$CDM cosmology mimics the $R_{\rm h}=ct$
Universe, from the present time, $t_0$, all the way back to when 
the Universe was only $t_0/7$ years old. The point is that with
all their complexity, the many free parameters in $\Lambda$CDM
must be adjusted to fit the data in such a way that the universal
expansion is essentially what it would have been anyway
in the $R_{\rm h}=ct$ Universe.}
\end{figure}

Several characteristics of this comparison are rather striking. First, the two curves are virtually
identical, and not only at low redshifts, but all the way out to $z=6$ or more. They differ slightly
around $z\sim 1-2$, which should not be surprising given that the two luminosity distances
(Equations~9 and 13) depend quite differently on the assumed parameters. Yet, with all
the possible deviations one might have expected between these two formulations with increasing
$z$, one instead sees that $\Lambda$CDM is forced to track $R_{\rm h}=ct$ over the entire
redshift range.

Second, a big deal is made of the fact that in $\Lambda$CDM, the cosmological expansion
switched from deceleration to acceleration at about half its current age. That in itself is
a rather strange coincidence. Do we really live at such a special time that we are privileged
to have seen the Universe decelerate for half of its existence and then accelerate for the
other half, but in such a perfect balance that the two effects exactly canceled out?
The reasonable answer to this question is clearly no. And we can see in Figure~4 what 
must be happening. Since the $R_{\rm h}=ct$ constraint is imposed on the Universe
by the Cosmological principle and the Weyl postulate, even in $\Lambda$CDM the
cosmological distance scale must comply globally with this condition. So if the imperfect
formulation of $d_L$ in Equation~9 first forces a decelerated expansion, a compensating
acceleration must take place later in order to reduce the overall cosmological expansion
back to $R_{\rm h}=ct$ as $t\rightarrow t_0$.

But if the $R_{\rm h}=ct$ curve should be the best fit to the data, why does $\Lambda$CDM
provide such a good fit to the Union2.1 sample, even at $z\sim 0.7$? We suggest that
the problem lies with the so-called ``nuisance" variables used to determine the
supernova luminosities. When $\Lambda$CDM is fit to the data, these variables are
optimized along with the other parameters defining the cosmological model (what
we called $\xi$ in \S~3). With so much flexibility ($\alpha$, $\beta$, $\delta$, and $M_B$),
the distance moduli cannot be determined without the assumption of an underlying
cosmology. And if $\chi_{\rm stat}^2$ is minimized by varying all of the parameters
simultaneously, it should not be surprising to see the data complying with the model.

A very important contribution to this discussion was made recently by Seikel \&
Schwarz (2009), who improved the analysis of Type Ia SN data by eliminating
some of the uncertainties arising from specific model fitting. Instead of performing
a routine $\chi^2_{\rm stat}$ fit using Equation~(2), they considered the redshift dependence
of the quantity $\Delta\mu_i$ averaged over redshift bins. Each $\Delta\mu_i$
is defined to be the difference between the measured distance modulus of event
$i$ and the value it would have had at that redshift if the luminosity distance
were calculated assuming no universal acceleration, i.e., if it were given by 
Equation~(13). 

Their goal was simply to ascertain whether or not the average $\langle\Delta\mu\rangle$
is greater than zero---the ``null" hypothesis. A positive value would be taken as
evidence of recent acceleration. By comparing these quantities at redshifts
greater than 0.1 to those below, their goal was to remove as many systematic
uncertainties as possible, while at the same time not restricting their analysis
to any particular model. The attractive feature of this approach is that one
needs only to demonstrate that $\langle\Delta\mu\rangle>0$ for acceleration 
to have occurred, without relying on fits with a particular cosmology.

Interestingly, their principal conclusion was that acceleration emerges only if
one includes the events at $z<0.1$. Quite significantly, the evidence for
acceleration dramatically (to use their own terminology) decreases if SNe
with $z<0.1$ are not used for the test. In their careful assessment of
this effect, they pointed out that a redshift of 0.1 corresponds approximately
to a distance of about 400 Mpc, the size of the largest observed structures
in the Universe. In other words, the evidence for acceleration emerges only
when one includes events associated with a length scale over which the
assumption of homogeneity and isotropy is probably not justified.

Having said this, it must be pointed out that although their approach
appears to be superior to the simple $\chi^2_{\rm stat}$ fitting given in Equation~(2),
it nonetheless also suffers from the problem we have been describing---the 
model-dependence of the reduced data. Their $\Delta\mu_i$ are the 
differences between the measured distance moduli $\mu_i$ and the
values these would have at the same redshift in a non-accelerated 
universe. Unfortunately, in order to measure $\mu_i$, one must 
pre-assume a cosmology. as described in \S~3, for otherwise
it is not possible to determine the 4 nuisance parameters. And when
one adopts a particular cosmology, optimizing its free parameters
via the fitting procedure, the nuisance parameters are themselves
optimized to comply with that particular cosmology. The values of
$\mu_i$ used by Seikel \& Schwarz (2009) were therefore not
model-independent, as they assumed. Instead, they had already
been optimized for $\Lambda$CDM, ensuring that they would show
at least some vestige of acceleration relative to the null case.

Their work is useful nonetheless because they were apparently able 
to at least eliminate some of the systematic errors, and in so doing,
demonstrate quite compellingly that the elimination of events associated
with the inhomogeneous portion of the local Universe greatly decreases
the evidence for acceleration. 

In any fitting procedure, whether it be a direct $\chi^2_{\rm stat}$ fit according
to Equation~(2), or a more sophisticated approach based on Seikel
\& Schwarz's null hypothesis, what is clearly needed is a careful 
reduction of the data using the very same cosmology being tested.
It is not correct to optimize the nuisance parameters using one
model and then use them to test the predictions of another.
It is our hope that such an exercise will be carried out soon 
for a direct comparison between $\Lambda$CDM and the
$R_{\rm h}=ct$ Universe.

\section{Conclusions}
The analysis of Type Ia supernova data over the past decade has been one of the
most notable success stories in cosmology. These are arguably the most reliable
standard candles we have to date and offer us an unparalleled opportunity of studying
the cosmological expansion, at least out to a redshift of $\sim 1.5$. The consensus
today appears to be that $\Lambda$CDM offers the best explanation for the 
luminosity-distance relationship seen in these events, and many believe that
what remains to be done is simply to refine the best-fit values of the parameters
that define this model.

One may therefore question the value of re-examining the merit of $\Lambda$CDM
in these studies, and it would be difficult to justify re-testing the basic theory, were
it not for the significant incompatibility emerging between the standard model and other
equally important observations, such as the CMB anisotropies mapped by the
Wilkinson Microwave Anisotropy Probe (Bennett et al. 2003). 
In this paper, we have documented several
compelling reasons for believing that $\Lambda$CDM does not provide an
accurate representation of the cosmological expansion, at least not at high
redshift ($z>>2$). It has therefore been essential to re-analyze the Type Ia 
supernova data in light of the cosmology (the $R_{\rm h}=ct$ Universe)
that best represents the Universe's dynamical evolution at early times.

By directly comparing the distance-relationship in $\Lambda$CDM with
that predicted by the $R_{\rm h}=ct$ Universe, and each with the
Union2.1 sample, we have demonstrated that the two theories produce 
virtually indistinguishable profiles. If these were not related, such
a close empirical kinship would suggest the emergence of yet another ``cosmic
coincidence" in $\Lambda$CDM.   Instead, the fact that the best fit
parameters compel $\Lambda$CDM to track $R_{\rm h}=ct$ so closely
suggests that the former cosmology is an approximation to the latter,
and relaxes to it as best as the formulation in terms of $\Omega_m$,
$\Omega_r$, $\Omega_\Lambda$, and $w$ allows it to. 

But because the relaxation is not perfect, $\Lambda$CDM accelerates
in certain regions and decelerates in others, though always precisely
balanced in order to guarantee a net zero acceleration over the
age of the Universe. Were it not for the constraint imposed on
the gravitational radius by the Cosmological principle and the 
Weyl postulate, this balancing act would be one more cosmic
coincidence, as would the fact that the switch from deceleration
to acceleration occurs at the mid-point of the Universe's existence,
endowing us with the remarkable privilege of viewing this 
transition at a very special time---one that occurs once, and once 
only, in the entire history of the Universe.

Through this exercise, we have also highlighted the fact that
the data themselves cannot be determined independently of
the assumed cosmological model. The fact that the supernova
luminosities must be evaluated by optimizing 4 parameters 
simultaneously with those in the adopted model means that
the resultant data are compliant to the implied cosmology.
Though this approach is highly unpalatable, there doesn't seem 
to be any reasonable alternative at this time. However,
the dependence of the data on the assumed model suggests
that one should not ignore the model-dependent data reduction
in any comparative studies between competing theories. 

It would therefore be highly desirable to carry out a full
fitting analysis of the Union2.1 sample using the $R_{\rm h}
=ct$ Universe, on par with the efforts already undertaken for
$\Lambda$CDM. Only then can we see whether the Type Ia
supernova data reveal a cosmological expansion at low
redshifts consistent with the dynamics implied by the CMB
anisotropies, or whether they do in fact demonstrate a
different behavior at early and late times in the Universe's
history.

\acknowledgments
This research was partially supported by ONR grant N00014-09-C-0032
at the University of Arizona. I am grateful to Amherst College
for its support through a John Woodruff Simpson Lectureship.


\begin{thebibliography}

\bibitem[]{B03}
Bennett, C. L. et al. 2003, ApJ Sup, 148, 97

\bibitem[]{C09}
Copi, C. J., Huterer, D., Schwarz, D. J. \& Starkman, G. D. 2009, MNRAS, 399, 295

\bibitem[]{D09}
Dawson, K. S. et al. 2009, AJ, 138, 1271

\bibitem[]{G98}
Garnavich, G. et al. 1998, ApJ, 509, 74

\bibitem[]{H09}
Hicken, M. et al. 2009, ApJ, 700, 331

\bibitem[]{H12}
Hlozek, R. et al. 2012, ApJ, 749, 90

\bibitem[]{K10}
Kelly, P. L., Hicken, M., Burke, D. L., Mandel, K. S. \& Kirshner, R. P. 2010, ApJ, 715, 743

\bibitem[]{K09}
Kowalski, M. et al. 2008, ApJ, 686, 749

\bibitem[]{K08}
Kuznetsova, N. et al. 2008, ApJ, 673, 981

\bibitem[]{L10}
Lampeitl, H. et al. 2010, ApJ, 722, 566

\bibitem[]{M07}
Melia, F. 2007, MNRAS, 382, 1917

\bibitem[]{M12a}
Melia, F. 2012a, Australian Physics, 49, 83

\bibitem[]{M12b}
Melia, F. 2012b, Phys. Rev. D, submitted

\bibitem[]{M12c}
Melia, F. 2012c, Phys. Rev. D, submitted

\bibitem[]{M12d}
Melia, F. 2012d, Phys. Rev. D, submitted

\bibitem[]{M12}
Melia, F. \& Shevchuk, A. 2012, MNRAS, 419, 2579

\bibitem[]{P98}
Perlmutter, S. et al. 1998, Nature, 391, 51

\bibitem[]{P99}
Perlmutter, S. et al. 1999, ApJ, 517, 565

\bibitem[]{R98}
Riess, A. G. et al. 1998, AJ, 116, 1009

\bibitem[]{R04}
Riess, A. G. et al. 2004, ApJ, 730, 119

\bibitem[]{S98}
Schmidt, B. P. et al. 1998, ApJ, 507, 46

\bibitem[]{S95}
Scott, D., Silk, J. \& Martin, W. 1995, Science, 268, 829

\bibitem[]{S09}
Seikel, M. \& Schwarz, D. J. 2009, JCAP, 02, 024

\bibitem[]{S10}
Sullivan, M. et al. 2010, MNRAS, 406, 782

\bibitem[]{S12}
Suzuki, N. et al. 2012, ApJ, 746, article id. 85

\bibitem[]{W11}
Watson, D. F., Berlind, A. A. \& Zentner, A. R. 2011, ApJ, 738, article id. 22

\end{thebibliography}
\end{document}